\documentclass[usegraphicx,useAMS,usenatbib,referee]{mn2e}

\title[UV Bright Globular Clusters in M87]
    {UV Bright Globular Clusters in M87: More Evidence for Super-helium-rich Stellar Populations?}

\author[S.~Kaviraj et al.]
{\Large S. Kaviraj$^{1}$, 
S. T. Sohn$^{2,3}$, R. W. O'Connell$^{3}$, S. -J. Yoon$^{4}$, Y. W.
Lee$^{4}$ and S. K.
Yi$^{4}$\thanks{Send offprint request to yi@yonsei.ac.kr}\\
$^{1}$ Department of Physics, University of Oxford, Keble Road,
Oxford OX1 3RH, UK\\
$^{2}$ Korea Astronomy and Space Science Institute, 61-1
Hwaam-Dong, Yuseong-Gu, Daejeon 305­348, Korea\\
$^{3}$ University of Virginia, Department of Astronomy,
Charlottesville, VA 22904, USA\\
$^{4}$ Yonsei University, Center for Space Astrophysics, Seoul
120­749, Korea}

\begin{document}

\date{Accepted in MNRAS: \today}

\pagerange{\pageref{firstpage}--\pageref{lastpage}} \pubyear{2007}

\maketitle

\label{firstpage}

\begin{abstract}
We study the $UV$ and optical properties of 38 massive globular
clusters (GCs) in the Virgo elliptical, M87, imaged using the
$STIS$ and $WFPC2$ instruments onboard the \emph{Hubble Space
Telescope}. The majority of these GCs appear
extremely bright in the far-ultraviolet ($FUV$) - roughly a
magnitude brighter than their Galactic counterparts with similar
metallicities. The observed $FUV$ flux is several times larger
than predictions of canonical old stellar population models.
% with \emph{physical} ages i.e. ages less than the age
%of the Universe. 
These canonical models, which assume a fiducial
helium enrichment parameter, $\Delta Y/\Delta Z=2$, are able to
reproduce the observed $FUV$ fluxes only if ages $\sim$3--5\, Gyr 
larger than the ``WMAP age'' of the Universe are invoked, 
although the same models fit the $UV$ photometry of Galactic and M31 GCs
for ages less than the ``WMAP age''. A similar discrepancy 
($\sim3$\,Gyr) is found between the mass-weighted and $UV$-luminosity
weighted ages of the massive Galactic GC $\omega$~Cen, whose
colour-magnitude diagram (including peculiar features on its
well-populated horizontal branch) can be accurately
reproduced by invoking a small super-He-rich ($\Delta Y/\Delta
Z \ga 90$) stellar component. 
By comparison to $\omega$~Cen, we propose
that the majority of M87 GCs in our sample contain strong
signatures of similarly minor super-He-rich sub-components.
This hypothesis is supported by simulations which suggest that, based
on the $UV$ detection limit of this survey, the number of GCs
detected is several times of the prediction from canonical models.
%what would be expected if
%the GC population conformed to reasonable ages (9-13 Gyr), with a
%uniform helium enrichment of $\sim$2. 
Although we cannot prove or disprove the extreme helium scenario 
at the moment, we show that the same phenomenon that causes the 
extended horizontal branch of $\omega$~Cen explains the UV brightness 
of our sample. If this is indeed due to the extreme helium,
this study would be the first to find its signatures in extragalactic objects.
\end{abstract}

%...................................................................................................


%...................................................................................................

\section{Introduction}
Recent observations of the stellar population of the largest
Galactic globular cluster (GC), $\omega$~Cen, have revealed a
double main sequence (MS), with a disjoint sub-population of bluer
and fainter MS stars which is separated from a majority population
of redder and brighter MS stars \citep[e.g.][]{Bedin2004}. In
addition, $\omega$~Cen harbours a substantial population of
extended horizontal branch (EHB) stars, which are hotter and
distinct from the primary population of normal horizontal branch
(HB) stars
\citep{Whitney1994,Ferraro2004,Sollima2005,Sollima2006}.
Theoretical analysis has indicated that the double main sequence
could be a result of a large spread in helium (He) abundances
within the cluster \citep{Norris2004} - this is supported by
spectroscopic measurements of metal abundances of MS stars in
$\omega$~Cen \citep{Piotto2005}. More recently, \citet{Lee2005}
have shown that a large variation in the He abundance in
$\omega$~Cen can not only reproduce the double MS, but also
accounts naturally for the `anomalous' population of hot extended
HB stars in this cluster\footnote{Note that the evidence for He
enrichment in $\omega$~Cen does not come from direct observations
of He spectral lines but is inferred from the HB morphology and
direct determinations of metallicity.}. Their best simulation of
$\omega$~Cen requires the  He enrichment ($\Delta Y/\Delta Z$)
parameter, for a minority population of stars, to be in excess of
90 - for comparison $\Delta Y/\Delta Z \sim 2$ on galactic scales
\citep[e.g.,][]{Peimbert2001}. Although such high values of helium
enrichment are difficult to reproduce theoretically
\citep[e.g.,][]{Bekki2006,Choi2007,Karakas2006}, 
similar super-He-rich sub-populations are
also capable of reproducing the peculiar HB morphology of another
Galactic GC, NGC 2808 \citep{Dantona2004,Bedin2000,Lee2005,Dantona2005}.
A similar phenomenon may be responsible for the unusually-blue
HB morphology of NGC\,6338, and NGC\,6441 for their metallicities   
\citep[e.g.,][]{Sweigart1998,Caloi2006}.

Helium-rich populations are thought to have a significantly higher
number of hot, evolved HB stars and thus show enhanced $UV$ fluxes
(and hence bluer integrated $UV$ colours) than helium-poor
counterparts of the same age \citep[e.g.][]{Dorman1995,Yi1997,Yi1998,Yi1999}.
Using models with standard He enrichment ($\Delta Y/\Delta Z
\sim 2$) to derive ages from such integrated $UV$ colours would
therefore lead to those ages being \emph{overestimated}, since
populations with standard He enrichment take \emph{longer} to
output the same level of $UV$ flux as those with a super-He-rich
sub-component. Reversing this argument implies that the detection
of GCs that are too blue to be produced by old populations with
standard He enrichment, might indicate the presence of anomalously
high He concentrations in the stellar populations within these
clusters. In this paper we present a plausibility argument for the
presence of just such a population of GCs in the elliptical galaxy
M87. The motivation for this work stems from the recent $UV$ study
of M87 globular clusters \citep[see][]{Sohn2006}. We find that (a)
the $FUV$ detection rate of M87 globular clusters in this study is
\emph{significantly} larger than can be expected from simulations
of a standard old stellar population that has a He enrichment
value of $\sim2$ and that (b) the majority of globular clusters
which \emph{are} detected have derived ($FUV$-weighted) ages which
are far in excess of the presently-accepted age of the Universe. 
In this study, we adopt the most recent estimate from the WMAP 
data as the ``physical'' age of the Universe \citep{Spergel2006}.
Since the stellar models used to perform the GC age estimation are 
well-calibrated to Galactic GCs, and produce physical ages 
($\la 13$ Gyr) for old GCs in M31, we argue that both pieces of 
evidence are consistent with the 
existence of super-He-rich populations in the GC system of M87,
and that indeed it may be the $FUV$-enhanced `super-helium-enriched'
tail of the GC distribution which has been sampled by this study,
given that the $FUV$ detection rate is far higher than that
expected for a normal population of HB stars.

%...................................................................................................

\section{Parameter estimation: age and metallicity of M87 GCs}
We begin by combining $HST/STIS$ broad-band $FUV$ and $NUV$
photometry from \citet{Sohn2006} with $HST/WFPC2$ optical
photometry from \citet{Jordan2002} in the $F336w$, $F410m$,
$F467m$ and $F547m$ filters and photometry in the $V$ and $I$
filters from \citet{Kundu1999}, to simultaneously estimate the
ages and metallicities of 38 GCs in M87. GC star formation
histories (SFHs) can be adequately parametrised by (dustless)
simple stellar populations (SSPs) of a given age and metallicity
alone, and our parameter estimation employs a finely-interpolated
grid of SSPs using the Yi-Yoon stellar models \citep[see
e.g.][]{Yi2003}. The parameter estimation proceeds as follows:

For a vector $\textbf{X}$ denoting parameters in the model, and a
vector $D$ denoting the measured observables (data),

\begin{equation}
\textnormal{prob}(\textbf{X}|D) \propto
\textnormal{prob}(D|\textbf{X}) \times
\textnormal{prob}(\textbf{X}),
\end{equation}

where $\textnormal{prob}(\textbf{X}|D)$ is the probability of the
model given the data (which is the object of interest),
$\textnormal{prob}(D|\textbf{X})$ is the probability of the data
given the model and $\textnormal{prob}(\textbf{X})$ is the prior
probability distribution of the model parameters

If we assume a uniform prior in our model parameters i.e.

\begin{equation}
\textnormal{prob}(\textbf{X})=\textnormal{constant},
\end{equation}

we have

\begin{equation}
\textnormal{prob}(\textbf{X}|D) \propto
\textnormal{prob}(D|\textbf{X}).
\end{equation}

Assuming Gaussian errors on the measured quantities gives

\begin{equation}
\textnormal{prob}(D|\textbf{X}) \propto \exp(-\chi^2/2),
\end{equation}

where $\exp(-\chi^2/2)$ is the likelihood function and $\chi^2$ is
the sum of the normalised residuals between the model-predicted
observables $M_k$ and the observed values $D_k$

\begin{equation}
\chi^2 = \sum^N_{k=1}\bigg(\frac{M_k-D_k}{\sigma_k}\bigg)^2,
\end{equation}

and $\sigma_k$ is the error in the residual $M_k-D_k$.
$\textnormal{prob}(\textbf{X}|D)$ is a \emph{joint} PDF, dependent
on all the model parameters. To isolate the effect of a single
parameter $X_1$ in, for example, a two parameter model
($\textnormal{prob}(\textbf{X}|D) \equiv
\textnormal{prob}(X_1,X_2|D)$) we can integrate out the effect of
$X_2$ to obtain the \emph{marginalised} PDF for $X_1$:

\begin{equation}
\textnormal{prob}(X_1|D)=\int^{\infty}_0
\textnormal{prob}(X_1,X_2|D)dX_2.
\end{equation}

In our analysis, the parameters $\textbf{X}$ in the model are the
age ($t$) and metallicity ($Z$) of the SSP in question, and the
observables are colours constructed from $UV$ and optical
photometry. We choose a uniform prior in $t$, in the range 1 to 18
Gyr. For the metallicity $Z$, we use a uniform prior in the range
$-2.3$ to 0.7 dex, which are essentially the metallicity limits of
the stellar models. For each GC analysed, we marginalise the age
and metallicity parameters, take the \emph{peak} ($x_0$) of the
marginal PDF $P(x)$ as the best estimate and \emph{define}
one-sigma limits $x_+$ and $x_-$ as follows:

\begin{eqnarray}
\int_{x_0}^{x_{+}} P(x)dx &=& 0.34 \int_{x_0}^\infty P(x)dx\\
\int_{x_{-}}^{x_0} P(x)dx &=& 0.34 \int_{-\infty}^{x_0} P(x)dx
\end{eqnarray}

We define the \emph{uncertainty} in the estimate of a parameter as
($x_{+} + x_{-}$) i.e. the total extent of the one-sigma limits in
parameter space.\\

\begin{figure}
\begin{center}
\includegraphics[width=5.5in]{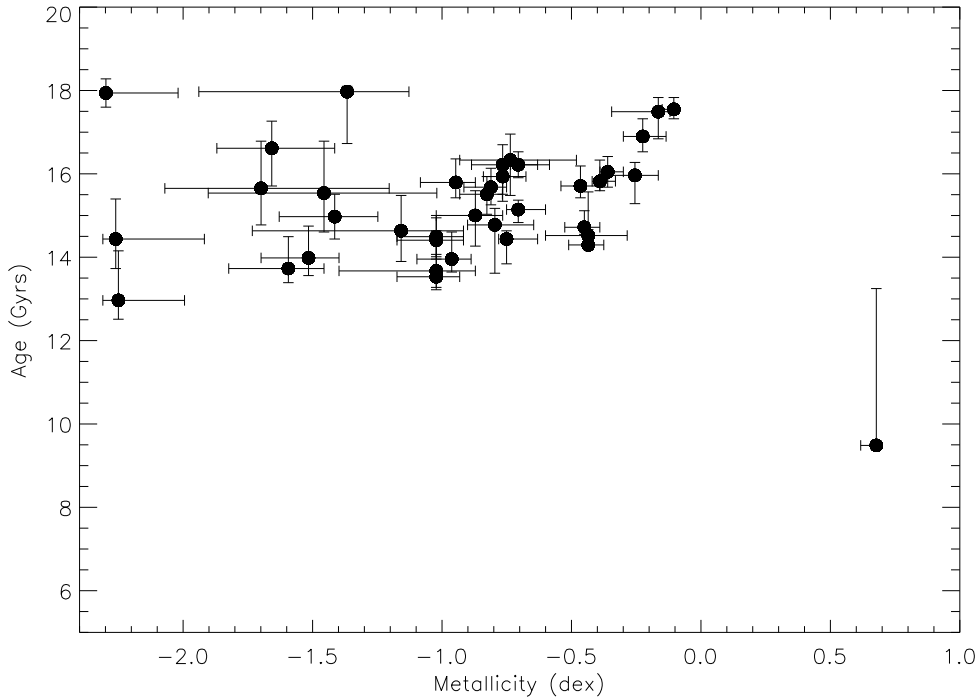}
\caption{Age and metallicity estimates for M87 GCs using $UV$ and
optical photometry (8 bands in total). \label{fig:f1} }
\end{center}
\end{figure}

In Figure \ref{fig:f1} we show the estimates of age and
metallicity for each GC, derived using this procedure from a set
of stellar models which has the fiducial value of He enrichment
($\Delta Y/\Delta Z=2$). We do not find a clear hint for a substantial
difference in age between the metal-poor and metal-rich GCs, 
in agreement with the result of \citet{Jordan2002}.

\begin{figure}
\begin{center}
\includegraphics[width=5.5in]{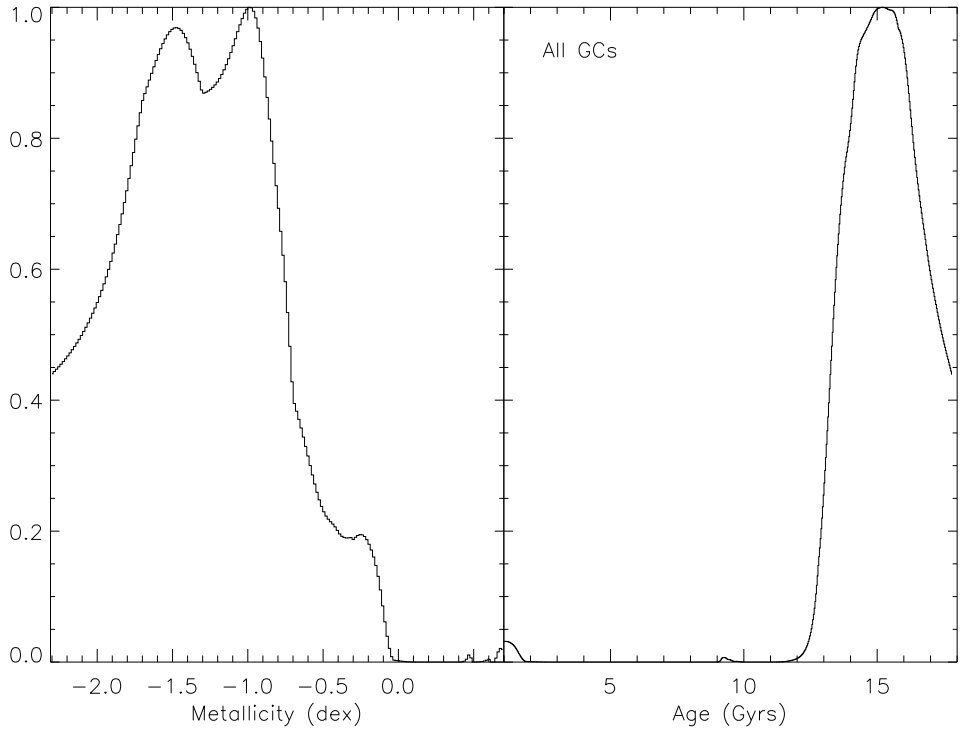}
\caption{Marginalised probability density functions for age and
metallicity for all M87 GCs used in this
study.}\label{fig:f2}
\end{center}
\end{figure}

Figure \ref{fig:f2} shows the marginalised
probability density function (PDF) for age and metallicity for all
M87 GCs used in this study. The metallicity PDF appears multi-modal.
Applying the KMM test for \emph{bimodality} yields two peaks: a
metal-poor peak at $-1.33$ dex and a metal-rich peak at $-$0.25 dex.
These values agree reasonably well with \citet{Jordan2002}, who
report values of $-$1.58 dex and $-$0.30 dex and \citet{Kundu1999},
who find values of $-$1.41 dex and $-$0.23 dex. The KMM test accepts
bimodality at the 98 percent confidence level and trimodality at
$>99.9$ percent confidence. However, we note that the trimodality,
in particular, is weak and that the small size of the sample,
coupled with its $UV$-selected nature makes it unreliable to
extrapolate this metallicity PDF to the entire M87 GC population!

Since the metallicity estimation in this study is derived from 8
photometric filters instead of one single colour index, it is
plausible that we are better able to resolve the metallicity of
the population and, in particular, the bimodality of the
metal-poor peak, which may not be obvious when using a single
optical colour such as $V-I$. We cannot, however, estimate the
\emph{proportion} of each subpopulation from this data, because
the GCs represent a \emph{FUV selected sample} and are therefore
unrepresentative of its parent population. 
%We should expect, for
%example, that the metal-rich peak is suppressed because metal-rich
%GCs would typically not be as $FUV$ bright as their metal-poor
%counterparts.

%...................................................................................................

\section{The derived ages of M87 globular clusters}
\begin{figure}
\begin{center}
\includegraphics[width=5.5in]{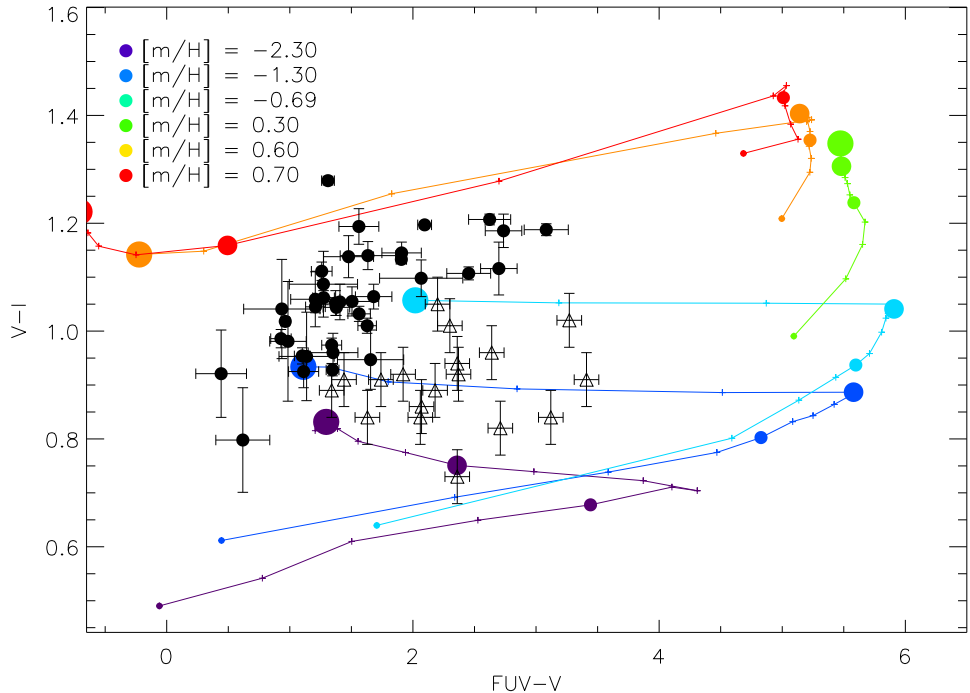}
\caption{Model $(FUV-V)$ vs. $(V-I)$ grid for a range of
metallicities and ages, generated from stellar models with the
fiducial value of He enrichment ($\Delta Y/\Delta Z=2$). The
lowest age plotted is 1 Gyr and the largest age plotted is 15
Gyr. Ages 1,5,10 and 15 Gyr are shown using filled circles of
increasing sizes. The GC data of M87 (filled circles) and Milky Way 
(open triangles) with errors are overplotted. 
It is apparent that the M87 photometry lies
\emph{outside} the age range 1-14 Gyr for all metallicities.}
\label{fig:f3}
\end{center}
\end{figure}

The ages derived in the last section, using the fiducial stellar
models, range 14--18 Gyr, unphysically large given current
estimates of the age of the universe. The technical reason for
this apparent `overestimate' of the age is that the M87 photometry
largely lies \emph{outside} the model colour-colour grids which
cover the age of the universe ($t\sim14$ Gyr). We illustrate
this by showing a model $(FUV-V)$ vs. $(V-I)$ grid in Figure
\ref{fig:f3}.
While the Milky Way GC data (open triangles) are reproduced by the grid, 
most of the M87 data (filled circles) lie outside the age range 1-14 Gyr 
for the generally-accepted metallicity range, [m/H]$ \la +0.2$ 
\citep{Cohen1998}.

\begin{figure}
\begin{center}
\includegraphics[width=5.5in]{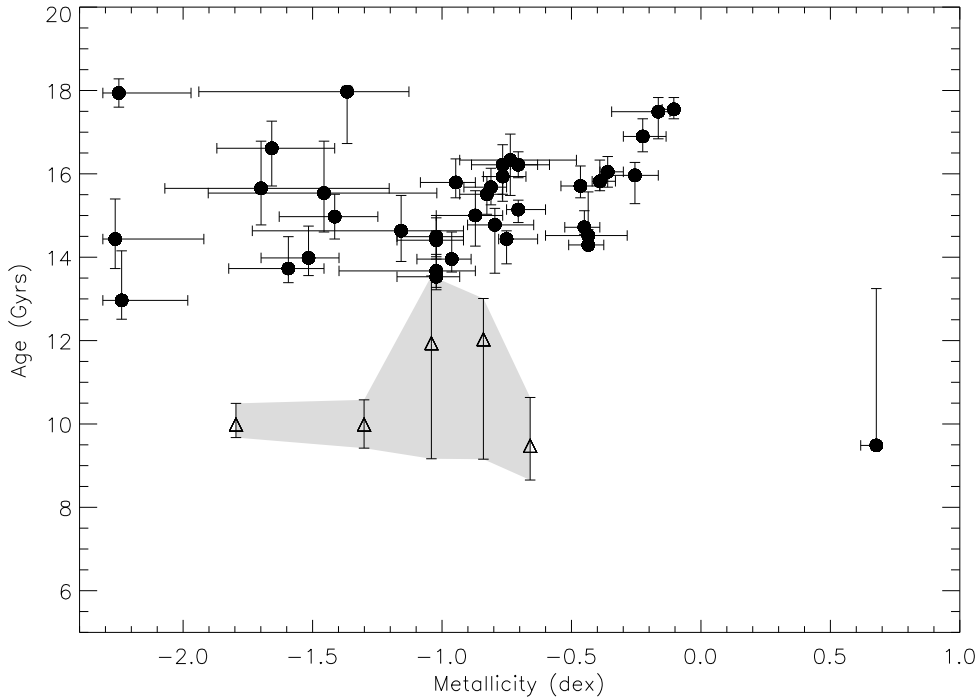}
\caption{Same as Figure \ref{fig:f1} but with the ages
of M87 GCs (studied by \citet{Cohen1998}) determined from Lick
absorption line indices and their uncertainties (open triangles
with error bars and grey shaded region) added. See Table 1 of
\citet{Cohen1998} for the Lick indices used in this age
determination.}\label{fig:f4}
\end{center}
\end{figure}

It is important to note that the ages derived for the M87 GCs in
the previous section are essentially \emph{FUV luminosity-weighted
ages}. For comparison, we now seek an estimate of the age of the
M87 GC population which \emph{does not} rely on their $FUV$ flux.
We note that \citet{Cohen1998} have previously derived ages of M87
GCs (as a function of metallicity) using the \citet{Worthey1994}
models. We repeat their analysis, using their measurements of the
Lick indices of M87 GCs (see their Table 1 for the specific
indices used), but this time using the \citet[][BC2003
hereafter]{BC2003} stellar models, which are more up to date. 
Using the BC2003 models, our Lick index-based age estimates on 
Cohen et al.'s M87 GCs are roughly 10--11\,Gyr, which is somewhat
lower than those of \citet{Cohen1998}.
Figure \ref{fig:f4} shows the same information as
Figure \ref{fig:f1}, with the ages determined from the
Lick indices overplotted using open triangles. \citet{Lee2000}
have previously noted that a population with a hot HB would appear
younger in Lick indices by $\sim 2$--3\,Gyr due to a significant
contribution from the HB stars to the $B$-band flux. Even considering
this effect, the Lick index ages are still 2--4\,Gyr younger
than the $UV$-derived ages.

We note here that the \citet{Cohen1998} sample does not overlap
with the \citet{Sohn2006} sample of GCs. Although we would
ideally like to determine the `spectroscopic' ages of the Sohn et
al. GCs, no absorption line data exists for this sample and
therefore we are not able to extract ages and metallicities for
these objects independent of their $UV$-optical colours. However,
if we now assume that (a) the \citet{Cohen1998} sample is
representative of the \emph{parent} M87 GC population and (b) the
age derivation from Lick indices is closer to the \emph{true} age
of the M87 GC population than that derived from the $UV$
photometry, then we find a $\sim 2-3$ Gyr discrepancy between the the
$FUV$ luminosity-weighted mean age of M87 GCs in this sample and
the `true' age of the parent population. The discrepancy is driven
by an enhancement of the $FUV$ flux in these clusters beyond that
expected from normal HB stars. 

%...................................................................................................

\section{$FUV$ enhancement in M87 GCs: super-He-rich stars?}
\citet{Lee2005} have recently suggested that a small super-He-rich
stellar component reproduces both the peculiar features of the
main sequence and the anomalous EHB population (which is hotter
and emits significantly more $FUV$ flux than normal HB stars) in
$\omega$~Cen. In this section, we explore the plausibility of a
similar mechanism being responsible for the enhanced $FUV$ flux
seen in M87 GCs.

In order to explore the effect of enhanced He enrichment on the
age derivation of the M87 GCs, we repeat the parameter estimation
of Section 2 but replace the fiducial stellar model used
previously ($\Delta Y/\Delta Z=2$) with a modestly helium-enhanced
model where $\Delta Y/\Delta Z=3$.

\begin{figure}
\begin{center}
\includegraphics[width=5.5in]{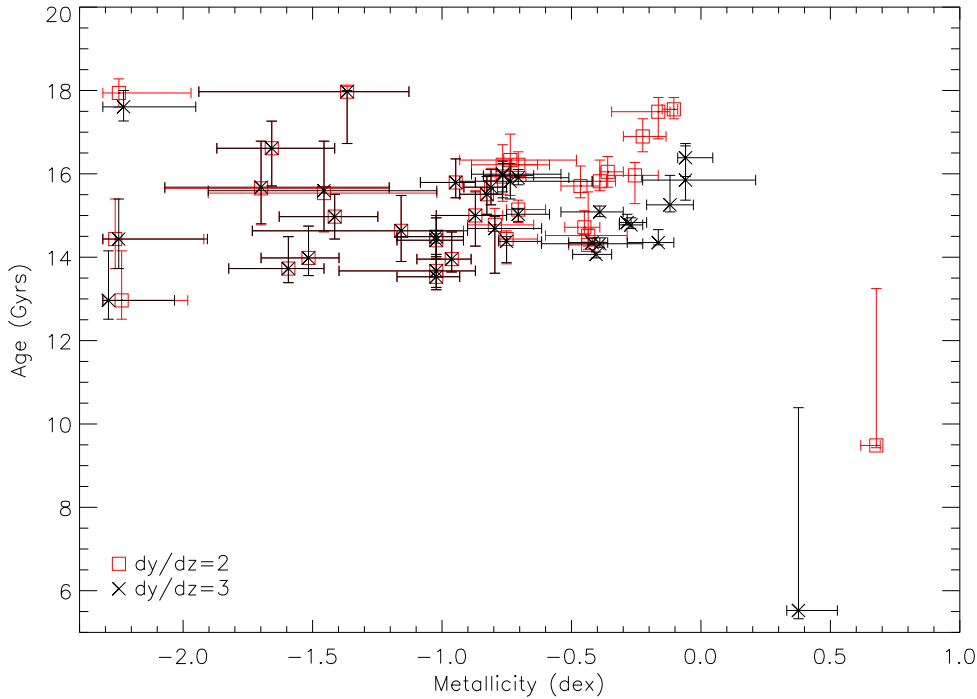}
\caption{Comparison between age and metallicity derivations using
the fiducial ($\Delta Y/\Delta Z=2$, open red squares) and helium-enhanced
($\Delta Y/\Delta Z=3$, black crosses) stellar
models.}\label{fig:f5}
\end{center}
\end{figure}

In Figure \ref{fig:f5} we compare the age and
metallicity derivations using the fiducial (open squares) and
helium-enhanced (crosses) stellar models. We find that GCs which have
best-fit metallicities higher than -1.0 dex show a \emph{lower}
best-fit age with the helium-enhanced model than the fiducial model.
The use of an helium-enhanced model moves the $FUV$ luminosity-weighted
ages of the GCs closer to their `true' ages, since He-rich stars
are hotter and thus brighter in the $UV$ than their He-poor counterparts
for the same age and mass. Nevertheless, the modest helium-enrichment
employed in the $\Delta Y/\Delta Z=3$ model does not lower the
estimated ages sufficiently to achieve agreement with the
spectroscopic ages.

The He enrichment ($\Delta Y/\Delta Z>90$), for a sub-population
of stars, adopted in \citet{Lee2005} is significantly more
extreme. To explore whether such extreme He enrichment could be
present in the M87 GCs, we compare the $FUV$ luminosity-weighted
age of $\omega$~Cen with its \emph{probable} true age. To compute
the true age of $\omega$~Cen, we calculate the \emph{mass-weighted}
age of the best model of \citet{Lee2005}, which accurately
reproduces the observed CMD of $\omega$ Cen (see Table 1 in their
paper). This mass-weighted age is $\sim 12$ Gyr. We then compute
the $FUV$ luminosity-weighted age for this GC, using its
$(15-V)_{0}$, $(25-V)_{0}$ and $(V-I)_{0}$ colours, taken from
\citet{Dorman1995}, and by performing the parameter estimation
with the fiducial $\Delta Y/\Delta Z=2$ stellar model.

\begin{figure}
\begin{center}
\includegraphics[width=5.5in]{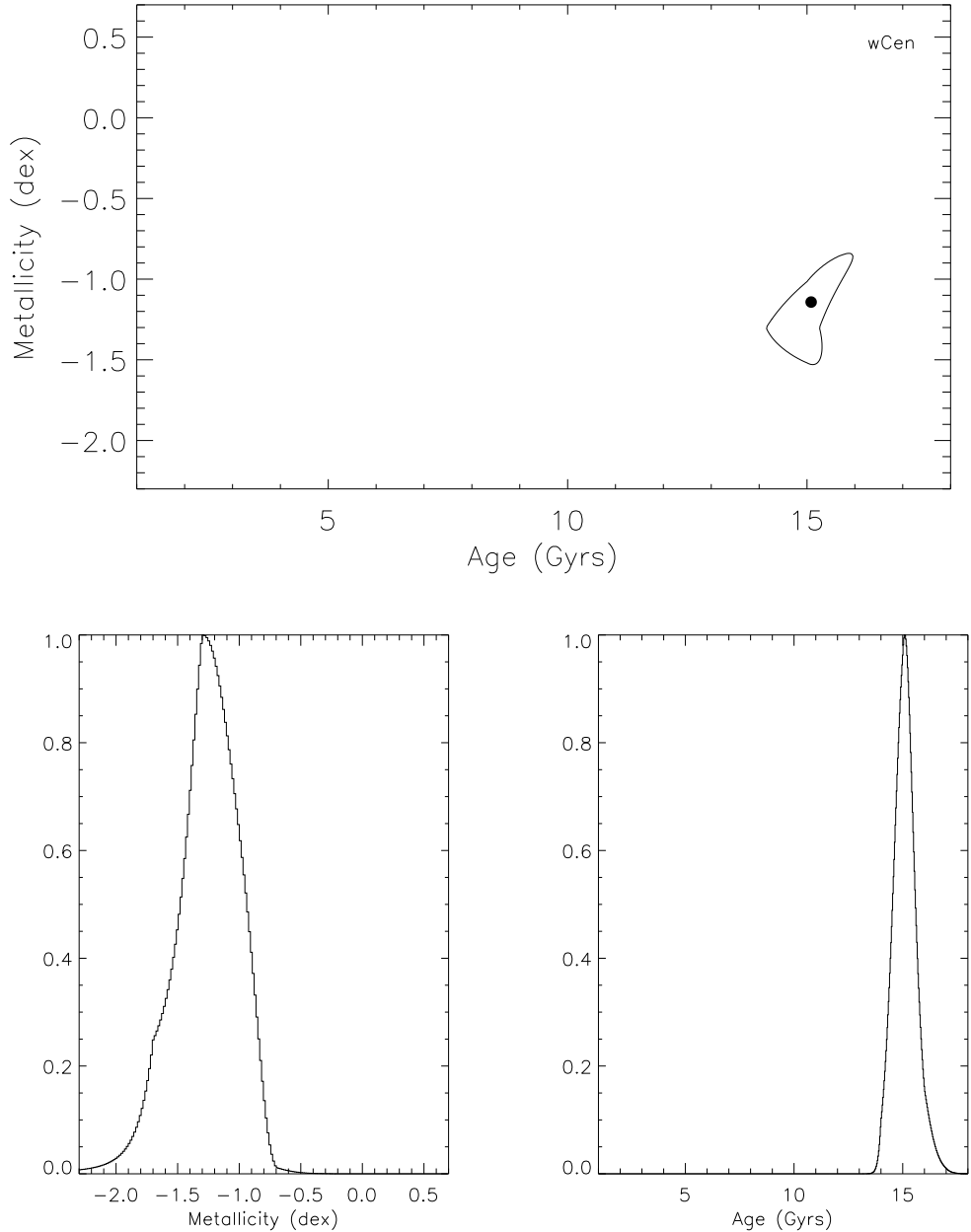}
\caption{TOP PANEL: Best-fit age and metallicity estimate for
$\omega$ Cen and one-sigma contour. BOTTOM LEFT: Marginalised
metallicity probability density function for $\omega$ Cen. BOTTOM
RIGHT: Marginalised age probability density function for $\omega$
Cen.}\label{fig:f6}
\end{center}
\end{figure}

As Figure \ref{fig:f6} indicates, the best-fit
metallicity for $\omega$~Cen is around $-1.3 \pm 0.21$ dex (for
comparison Dorman et al., 1995 quote the metallicity of
$\omega$~Cen as -1.6 dex) and the best-fit age is $15.08 \pm 0.4$
Gyr. We find that the discrepancy between the $FUV$
luminosity-weighted age and the mass-weighted age of $\omega$~Cen
is $\sim$ 3 Gyr, which is very similar to what we found for the
M87 GC population detected in the \citet{Sohn2006} study.

\begin{figure}
\begin{center}
\includegraphics[width=5.5in]{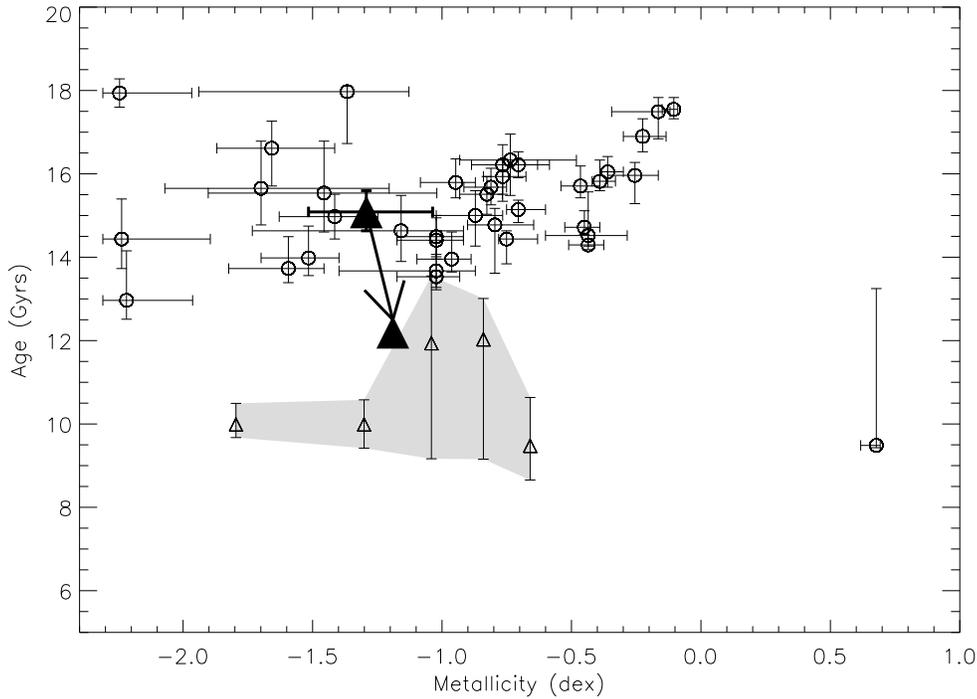}
\caption{Age and metallicity estimates for M87 GCs, derived using
$UV$ and optical photometry (black circles with error bars). Ages
of M87 GCs studied by \citet{Cohen1998}, determined from Lick
indices, are shown using the open triangles and grey shaded
region. The $FUV$ luminosity-weighted age (15.1 Gyr) and
mass-weighted age ($\sim$12 Gyr) of $\omega$~Cen are shown using
the large filled triangles.}\label{fig:f7}
\end{center}
\end{figure}

We collate all our findings in Figure
\ref{fig:f7}. The filled circles show age and
metallicity estimates for M87 GCs, derived using the $\Delta
Y/\Delta Z=2$ model. Ages for M87 GCs, determined from the Lick
indices, are overplotted using open triangles. The $FUV$
luminosity-weighted (15.1 Gyr) and mass-weighted (12 Gyr) ages
of $\omega$~Cen are shown by the large filled triangles. The
striking correspondence between the discrepancies in the
luminosity and mass-weighted ages in both $\omega$~Cen and the M87
GC population, indicates that a similar He enrichment to what is
conjectured in $\omega$~Cen would produce a similar enhancement of
$FUV$ flux in M87 GCs, causing their ages to be similarly
overestimated by $\sim 3$ Gyr.

%...................................................................................................

\section{Simulating the M87 GC population}

Figure \ref{fig:f3} shows the observed $UV$ GC data from
M87 and the Milky Way. We find that even optically-red GCs
show strong $UV$ fluxes and blue $UV-V$ colours ($FUV-V\leq2$). 
\citet{Sohn2006} found that all optically bright ($V<22$) GCs in
M87 are also $UV$ bright, regardless of their metallicities. Both
in the Milky Way and M87 samples, there appears to a be a positive
and thus sensible correlation between $V-I$ and $FUV-V$; but the
M87 GCs are brighter in the $UV$ than their Milky Way counterparts
by at least a magnitude or more. We note that, for a sample of M31 GCs 
detected in the $FUV$ by the GALEX satellite mission, 
the same stellar models yield best-fit ages of 10-12 Gyr, using an 
identical parameter estimation technique (S. -C. Rey, private 
communication).

\begin{figure}
\begin{center}
$\begin{array}{c}
\includegraphics[width=5.5in]{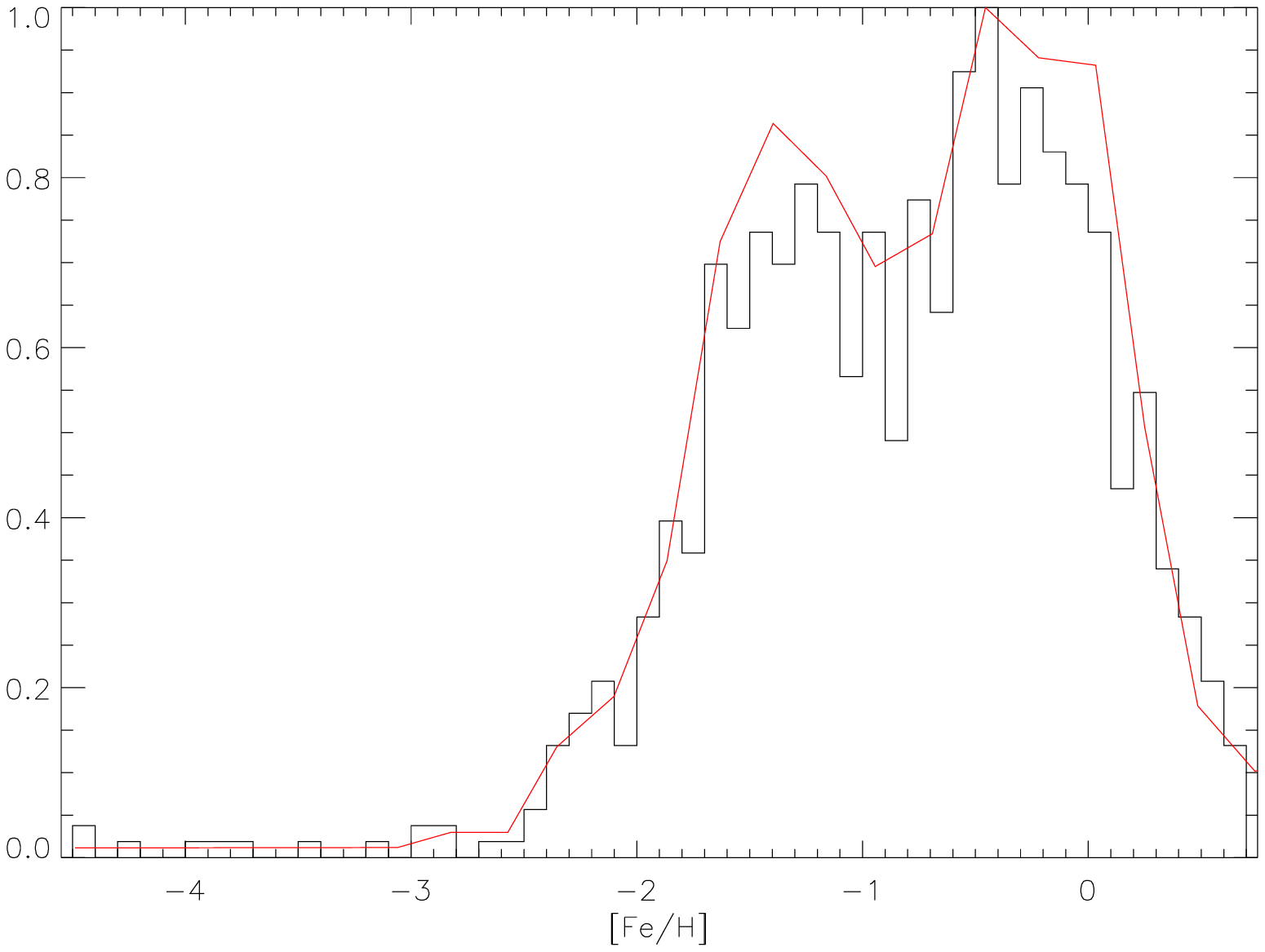}\\
\includegraphics[width=5.5in]{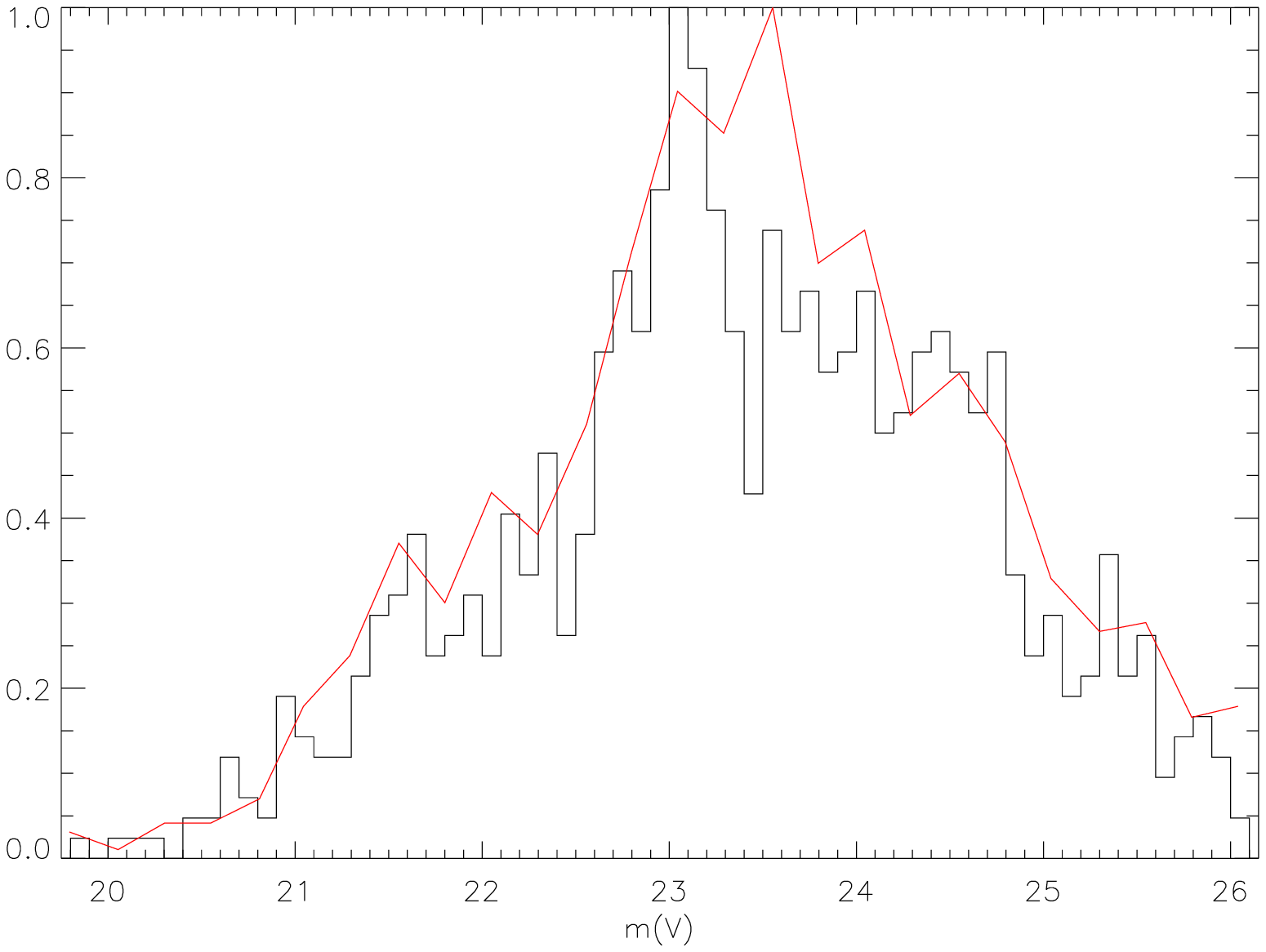}
\end{array}$
\caption{TOP: Simulated GC metallicities (histogram) and the
metallicity distribution function from \citet{Kundu1999} (solid
line). BOTTOM: Simulated GC $V$-band magnitudes (histogram) with
the observed GCLF from \citet{Kundu1999} (solid line). The plots
show a typical simulation of $\sim$1000 objects.}
\label{fig:f8}
\end{center}
\end{figure}

\begin{figure}
\begin{center}
$\begin{array}{c}
\includegraphics[width=5.5in]{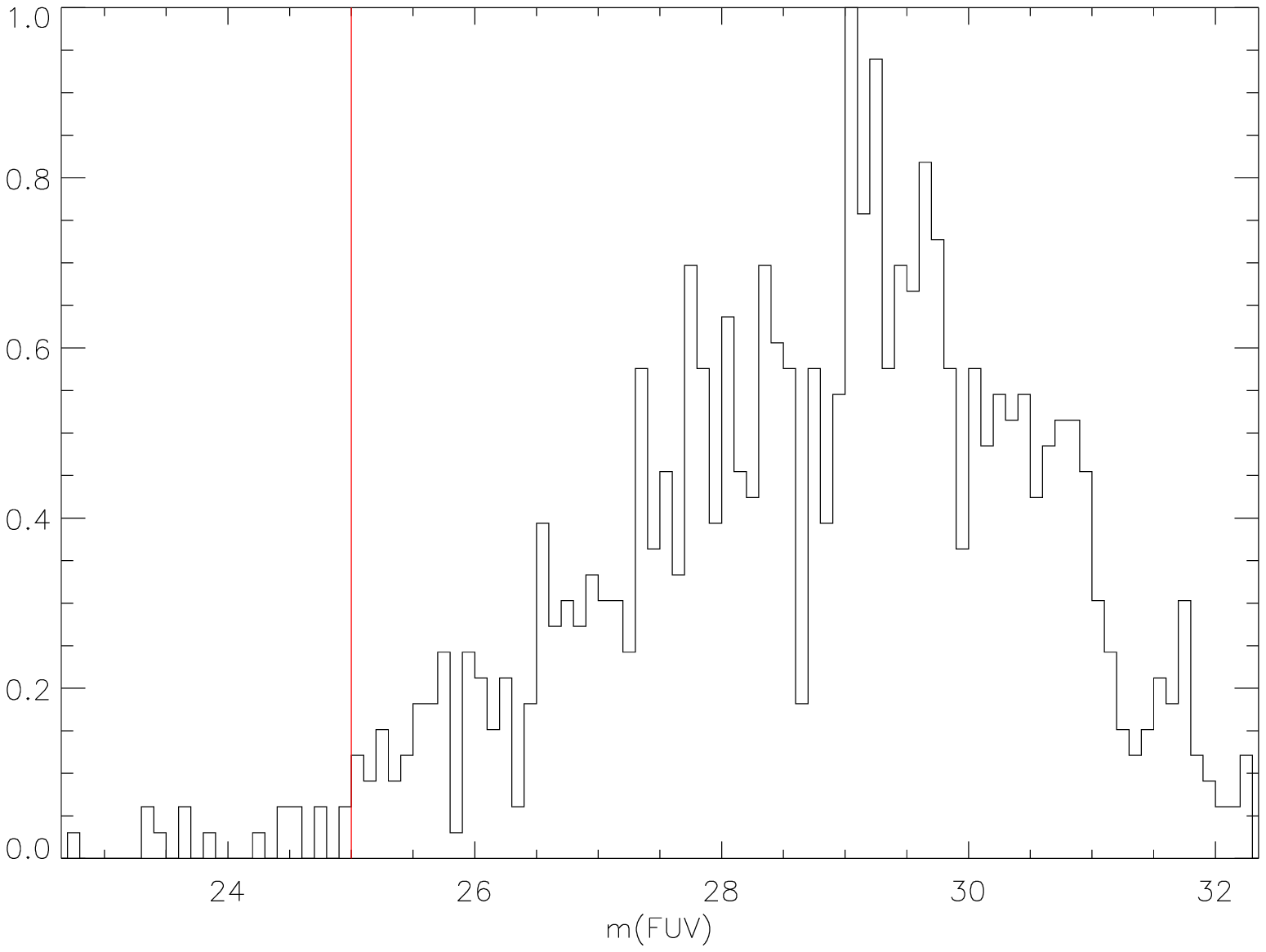}\\
\includegraphics[width=5.5in]{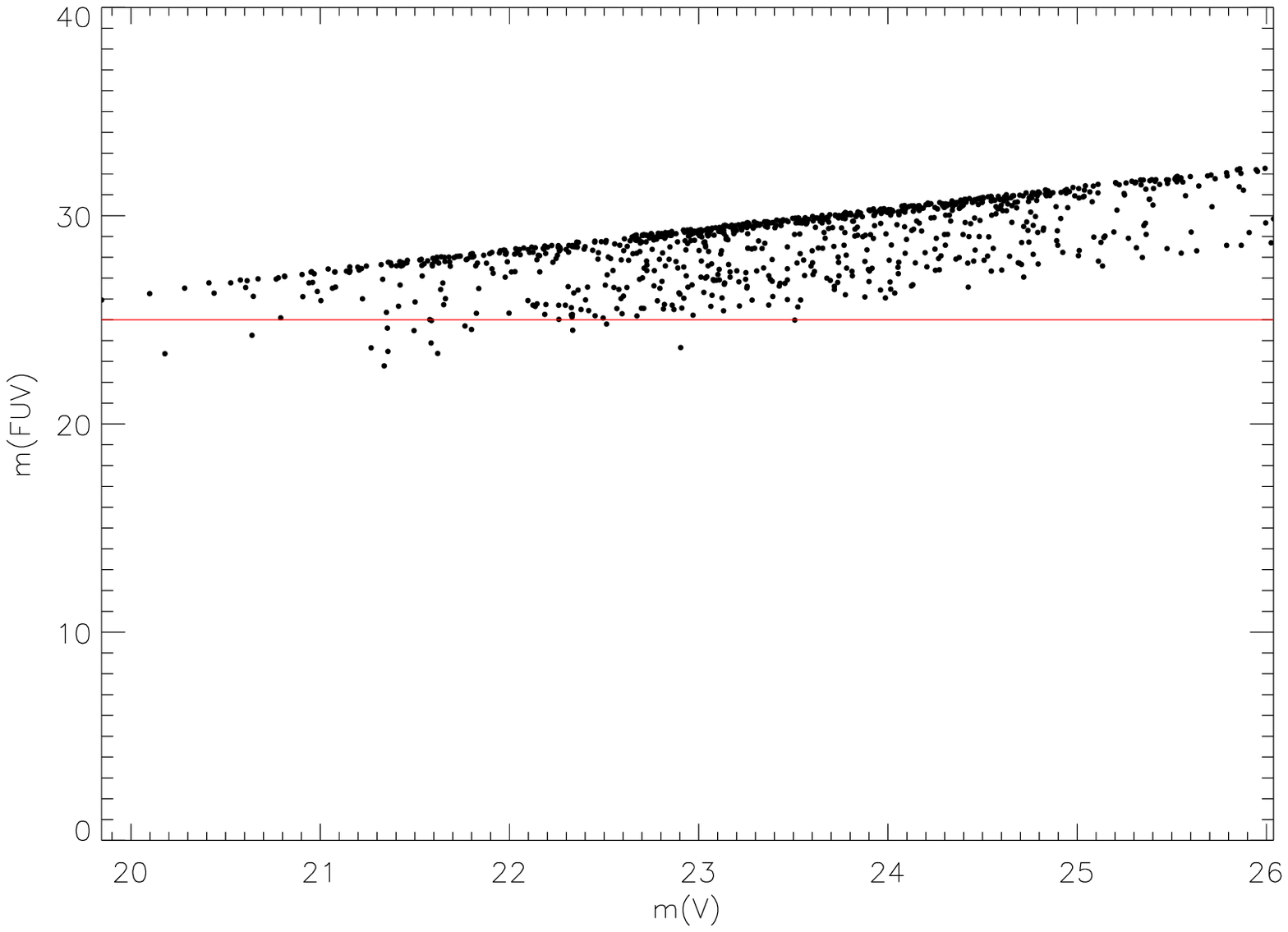}
\end{array}$
\caption{TOP: Simulated distribution of apparent GC $FUV$
magnitudes. The solid line shows the detection limit of the
\citet{Sohn2006} study. BOTTOM: Apparent $FUV$ magnitudes vs.
apparent $V$-band magnitudes for the simulated GCs. The solid line
shows the detection limit of the \citet{Sohn2006} study. The plots
show a typical simulation of $\sim$1000 objects.}
\label{fig:f9}
\end{center}
\end{figure}

To investigate possible reasons for these surprising features of
the M87 sample, we simulate a population of GCs using the GCLF and
GC metallicity distribution function (MDF) in M87 (Figure
\ref{fig:f8}), taken from \citet{Kundu1999}. Each GC is
allocated a formation age, drawn from a uniform distribution of
values between 9 and 13 Gyr. This yields a synthetic population of GCs,
each described by a $V$-band magnitude, a metallicity and a
formation age. Using the fiducial stellar models ($\Delta Y/\Delta
Z=2$), we then compute the apparent $FUV$ magnitude of each GC,
given its apparent $V$-band magnitude, metallicity and age. This
results in the distribution of apparent $FUV$ magnitudes  shown in
Figure \ref{fig:f9}. We now explore the characteristics
of \emph{GCs that we would expect to detect} in this simulated
sample, given the $FUV$ detection limit of 25 mag in the
\citet{Sohn2006} study. Note that, although we have performed
multiple ($\sim$50,000) simulations, the plots in this section
describe results from a typical simulation of $\sim$1000 objects.

Based on a large number of simulations, we find that the
\emph{detectable} fraction of \emph{normal} old GCs, is $2.1 \pm
0.6$ percent. We note that this value is reasonably robust
against the variation in the input distributions.
Since the \citet{Sohn2006} study (excluding Field 4
because it has a lower detection limit and is not representative
of the other fields) covers $\sim30$ percent of the WFPC2 image of
\citet{Kundu1999}, we would expect it to have $\sim300$ GCs in its
field of view. In an ideal scenario, where \emph{all} the
detectable GCs are indeed picked up by the observations, a
detection rate of $\sim$2 percent translates to $\sim$6 clusters
detected in the $FUV$. The \emph{actual} detection rate is
\emph{significantly higher}, with 66 sources detected by
\citet{Sohn2006}, of which 50 are confirmed as GCs through
comparison with the \citet{Kundu1999} study. The number is lowered
to 38 after cross-matching with \citet{Jordan2002}. We therefore
find that the detection rate achieved by this study is up to eight 
times higher than what might be expected on the basis of canoncial stellar
model predictions.  Given the expected detection rate of $\sim 2$\%, 
the observed detection rate of $\sim$17\% (50 out of 300)
achieved by this study is a 24 sigma event! 
%Given the extremely low
%expected detection rate for \emph{normal} old GCs, the lack of
%such GCs and the presence of enhanced $FUV$ fluxes in those GCs
%which \emph{are} detected, is therefore consistent with the
%results of this simulation.

\begin{figure}
\begin{center}
$\begin{array}{c}
\includegraphics[width=5.5in]{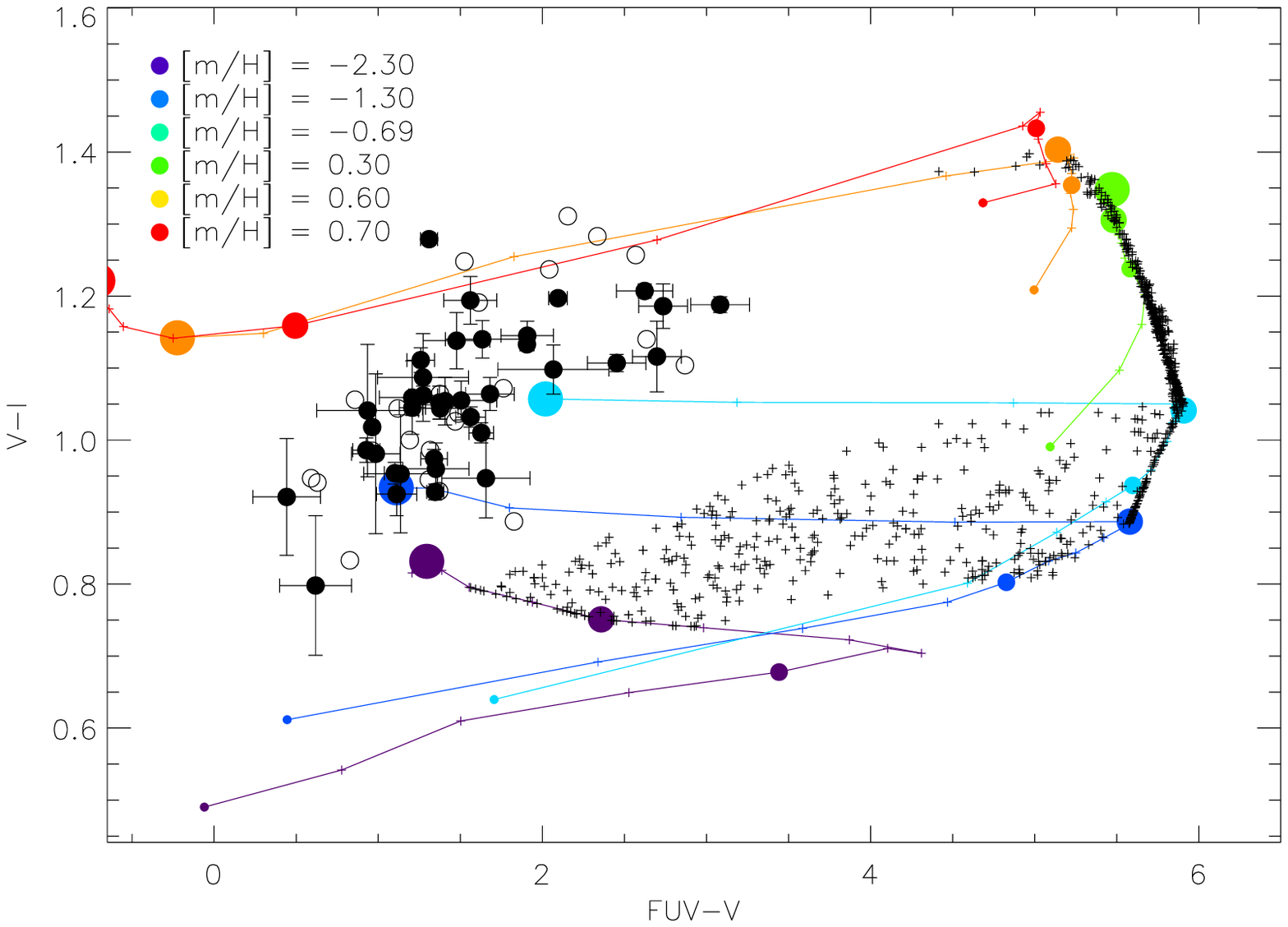}\\
\includegraphics[width=5.5in]{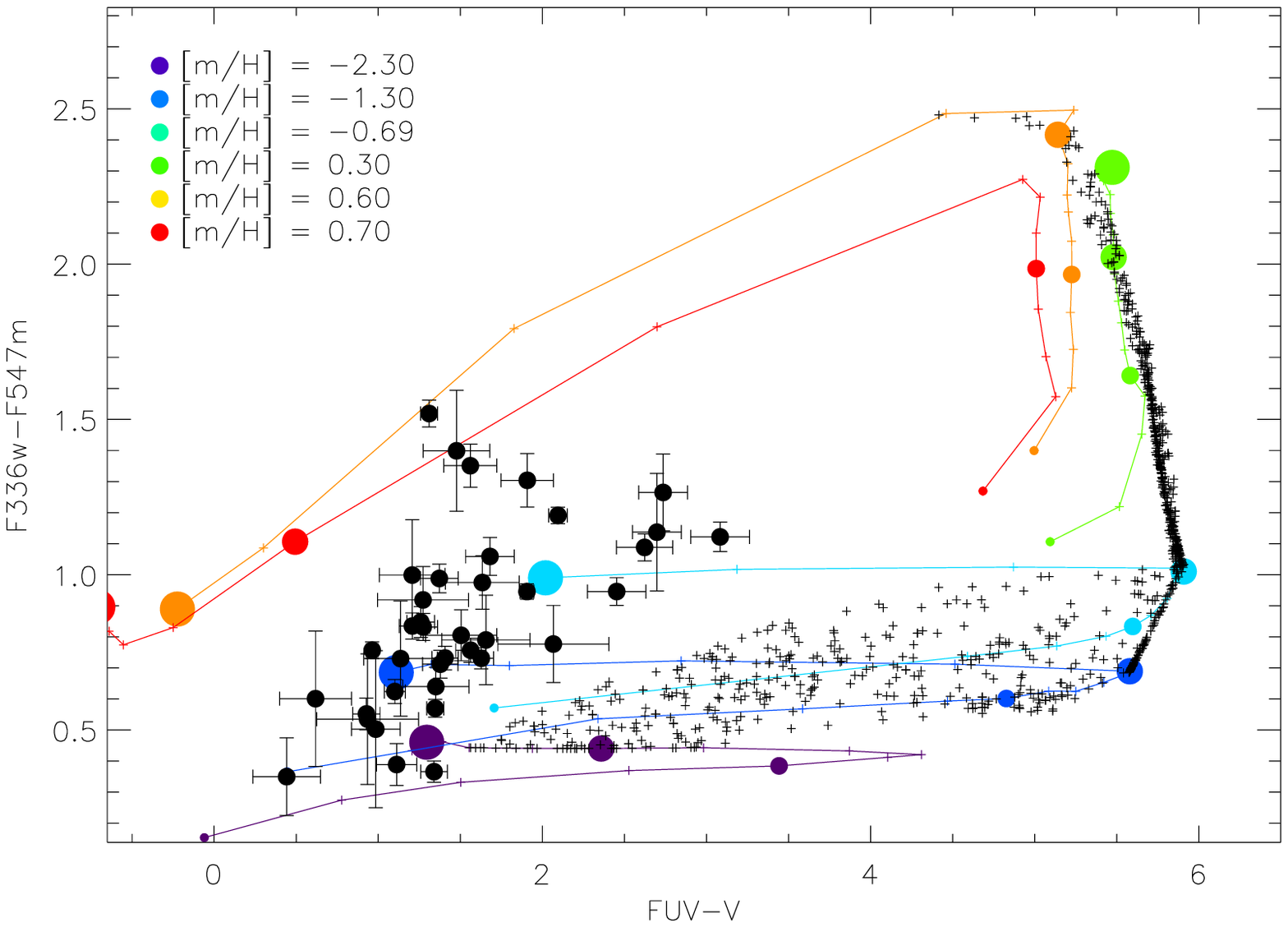}
\end{array}$
\caption{TOP: $(FUV-V)$ vs. $(V-I)$ for observed (filled circles
with error bars) and simulated GCs (crosses) plotted on top of a
model grid generated from stellar models with the fiducial value
of He enrichment ($\Delta Y/\Delta Z=2$). Filled circles represent
the 38 GCs used in this study after cross-matching the
\citet{Sohn2006} catalog with that of both \citet{Jordan2002} and
\citet{Kundu1999}. We also indicate, using open circles, GCs which
appear in \citet{Sohn2006} and \citet{Kundu1999} but not in
\citet{Jordan2002}, and which are therefore excluded from the
parameter estimation. The lowest model age plotted is 1 Gyr and
the largest age plotted is 15 Gyr. Ages 1,5,10 and 15 Gyr are
shown using filled circles of increasing sizes. BOTTOM: $(FUV-V)$
vs. $(F336w-F547m)$ for observed (filled circles with error bars)
and simulated GCs (crosses), plotted on top of a model grid
generated from stellar models with the fiducial value of He
enrichment ($\Delta Y/\Delta Z=2$).} \label{fig:f10}
\end{center}
\end{figure}

We finish this section by comparing the $FUV$ and optical colours
of the observed sample to those in the simulations (Figure
\ref{fig:f10}). As expected from the analysis presented in
the preceding sections, we see a clear separation between the
parameter space occupied by the simulated GCs and that occupied by
the observed sample - we illustrate this using grids which plot the
\citet{Sohn2006} $FUV$ photometry against optical photometry from
both \citet{Kundu1999} and \citet{Jordan2002}.

%...................................................................................................

\section{Summary and discussion}

We have studied the $UV$ photometry from $HST/STIS$ and optical
photometry from $HST/WFPC2$ photometry of 38 GCs in the Virgo
elliptical M87. 
%The strength of this (combined) dataset lies
%primarily in its homogeneity, and it 
The dataset represents the best photometric GC dataset covering 
both the $UV$ and the optical ranges. 
Parameter estimation of the age and metallicity
of each M87 GC, using stellar models with standard He enrichment
($\Delta Y/\Delta Z \sim 2$), results in almost all GCs having a
best-fit age of 14--18\,Gyr, significantly in excess of
the currently accepted age of the Universe.

The metallicity estimation leads to an age distribution which shows
no clear trend with metallicity at least in our limited sample. 
The marginalised metallicity PDF for the entire sample appears multi-modal. 
A KMM analysis for bimodality yields two peaks at -1.33
and -0.25 dex. These values agree (within errors) with previously
determined values in \citet{Jordan2002} and \citet{Kundu1999}. 
The significance of the agreement is however limited because our sample is a
$FUV$-selected subset of the parent GC population and thus are
affected by selection effects (and also low number statistics).
The derived ages of the GCs in this study are significantly larger
than the Lick-index-based ages of a more
representative sample of the M87 \emph{parent} GC population.
%taken from the study of \citet{Cohen1998}.

We have presented a \emph{plausibility argument} that the
`overestimate' in the derived ages may be driven by enhanced $FUV$
fluxes from anomalous HB components in these GCs, similar to those
postulated by \citet{Lee2005} in the Galactic GC $\omega$~Cen,
where a sub-population of EHB stars, which are hotter and
super-helium-enriched compared to the primary population of normal HB
stars, enhance the overall $FUV$ flux more than would be normally
expected.

To support this hypothesis, we have first shown that replacing the
fiducial stellar model, which assumes $\Delta Y/\Delta Z=2$, with
a modestly helium-enhanced stellar model ($\Delta Y/\Delta Z=3$)
lowers the best-fit ages of only the \emph{metal-rich} M87 GCs by
$\sim$ 1 to 2 Gyr. 
%We find, however, that such a modest change in
%the primordial He abundance is unable to reconcile the $FUV$
%luminosity-weighted and spectroscopically determined ages for the M87
%GCs in this study. 
We have then compared the $FUV$
luminosity-weighted age of $\omega$~Cen, derived from its $UV$
photometry in \citet{Dorman1995}, to its `true' age, calculated
from the mass-weighted mean age of the best-fit $\omega$~Cen model
of \citet{Lee2005}. We have found that this ``true'' age is around
12 Gyr compared to a $FUV$ luminosity-weighted age of around
15.1 Gyr - a discrepancy of ~3 Gyr, which is very similar to the
case of the $FUV$ bright M87 GCs in our sample.

Given the striking similarity in the discrepancy between the $FUV$
luminosity-weighted and mass-weighted ages of both $\omega$~Cen and
M87 GCs, we suggest that a similar type of extreme He
enrichment scenario could, 
in principle, produce the enhanced EHB components in the M87 GCs.
Given the fact that our M87 sample is $FUV$ selected, it is not
surprising that we detect mainly those GCs which show enhanced EHB
components (and therefore enhanced $FUV$ fluxes), leading to
overestimates of their ages when stellar models with fiducial
values of primordial He abundance are applied. This hypothesis is
supported by a simulation of M87 GCs, which
indicates that the detection rate for old GCs, given the GCLF, GC
MDF, area covered by the \citet{Sohn2006} study, and assuming
formation ages between 9 and 13 Gyr, should only be around 2
percent - the actual detection rate is up to eight times
higher. Since few GCs are expected to be observed
but a significant number \emph{are} detected, it is natural to
conclude that the $FUV$ fluxes are enhanced in the majority of
the detected GCs. 
Although we cannot prove its valitidy based on this argument, the 
super-He-rich hypothesis offers a reasonable explanation.

It is clear that drawing conclusions about the \emph{global} $FUV$
properties of the M87 GC population, based on this
``super-helium-enriched'' subsample alone, is not practical. Deeper
observations are required to sample both more GCs with enhanced
EHB components and also normal GCs which conform to the
expectations of the stellar models for old populations (such as
those found in M31). The presence of super-He-rich stellar
populations in Galactic GCs such as $\omega$~Cen and in the M87 GC
systems raises the intriguing possibility that this phenomenon may
be present, in varying degrees, in most systems in the nearby
Universe. A good example of such a case could be the UV upturn
phenomenon of elliptical galaxies
\citep{Code1979,Burstein1988,O'Connell1999}. 
Indeed, this study may be the first to discover signatures of apparent He
enrichment in extragalactic objects. Future $UV$ surveys of
nearby GC systems are therefore highly anticipated, both to
establish the possible ubiquity of the He enrichment phenomenon
and to measure the degree to which stellar populations may be He
enriched.

\section*{Acknowledgments}

We warmly thank Andres J\'{o}rdan for providing HST WFPC2 data on the
M87 globular clusters and for numerous comments, without which
this paper would not have been possible. He was involved in this work
from the beginning but declined our offer of the much-deserved co-authorship
following his work ethics, which we admire. We also thank the anonymous
refereee for encouragements and clarifications.
This work was supported by grant No. R01-2006-000-10716-0 from the Basic
Research Program of the Korea Science \& Engineering Foundation
and Yonsei University Research Fund to S.K.Y.

\bibliographystyle{mn2e}
\bibliography{references}

\end{document}